\ifx\texorpdfstring\undefined\newcommand\texorpdfstring[2]{{#1}}\fi
\documentclass[aps,prl,10pt,twocolumn,superscriptaddress,preprintnumbers,floatfix,nofootinbib,notitlepage,showkeys,showpacs]{revtex4-2}
\usepackage{multirow}
\usepackage{siunitx}
\usepackage{bbold}
\usepackage{braket}

\usepackage{microtype}
\usepackage[
  colorlinks,
  linkcolor = blue!60!black,
  urlcolor = blue!60!black,
  citecolor = blue!60!black,
]{hyperref}

\usepackage{tikz}
\usetikzlibrary{decorations.markings, arrows}

\usepackage{graphicx,times}
\usepackage{latexsym}
\usepackage{mathtools}

\usepackage{amsmath,amssymb,amsbsy,amsfonts}
\usepackage{array}
\usepackage{bm}
\usepackage{graphics}
\usepackage{mathrsfs}
\usepackage{xcolor}
\usepackage{cancel}
\usepackage[normalem]{ulem}

\usepackage{hyperref}
\usepackage[capitalise]{cleveref}

\usepackage{xcolor}
\usepackage{makecell}

\begin{document}

\title{Qubit Regularization of Asymptotic Freedom}
\preprint{LA-UR-20-29558}
\author{Tanmoy Bhattacharya}
\email{tanmoy@lanl.gov}
\affiliation{Los Alamos National Laboratory, Los Alamos, New Mexico 87545, USA}
\author{Alexander J. Buser}
\email{alexbuser@caltech.edu}
\affiliation{Institute for Quantum Information and Matter, Caltech, Pasadena, California 91106, USA}
\affiliation{Los Alamos National Laboratory, Los Alamos, New Mexico 87545, USA}
\author{Shailesh Chandrasekharan}
\email{sch27@duke.edu}
\affiliation{Department of Physics, Box 90305, Duke University, Durham, North Carolina 27708, USA}
\author{Rajan Gupta}
\email{rg@lanl.gov}
\affiliation{Los Alamos National Laboratory, Los Alamos, New Mexico 87545, USA}
\author{Hersh Singh}
\email{hershsg@uw.edu}
\affiliation{Department of Physics, Box 90305, Duke University, Durham, North Carolina 27708, USA}
\affiliation{Institute for Nuclear Theory, University of Washington, Seattle, Washington 98195-1550, USA}

\begin{abstract}
We provide strong evidence that the asymptotically free $(1+1)$-dimensional nonlinear $O(3)$ sigma model can be regularized using a quantum lattice Hamiltonian, referred to as the  ``Heisenberg comb'', that acts on a Hilbert space with only two qubits per spatial lattice site. The Heisenberg comb consists of a spin-half antiferromagnetic Heisenberg-chain coupled antiferromagnetically to a second local spin-half particle at every lattice site. Using a world-line Monte Carlo method, we show that the model reproduces the universal step-scaling function of the traditional model up to correlation lengths of $\num{200000}$ in lattice units and argue how the continuum limit could emerge. We provide a quantum circuit description of time evolution of the model and argue that near-term quantum computers may suffice to demonstrate asymptotic freedom.
\end{abstract}

\maketitle

Formulating quantum field theories (QFTs) so that they can be implemented on a quantum computer has become an active area of research recently~\cite{Jordan:2011ne,Casanova:2011wh,Casanova:2012zz,PhysRevX.6.031007,Klco:2018kyo,Klco:2019evd,Alexandru:2019nsa,Banuls:2019bmf,Banuls:2019rao,Bender:2020jgr}. One of the first steps in this process is to construct a suitable lattice quantum Hamiltonian that acts on a Hilbert space realized by $n$ qubits at each lattice site where $n$ is small. For bosonic quantum field theories, including gauge theories, an exact realization of the canonical commutation relation 
$ [\phi_x,\pi_y] = i \delta_{x,y}$ forces $n$ to be infinite. 
For this reason, all traditional formulations of lattice QFTs with bosonic degrees of freedom need reformulation to be solvable on a digital quantum computer. A simple way to proceed is to truncate the infinite dimensional Hilbert space to an $n$-qubit subspace while preserving the long distance physics. Universality suggests that long distance physics can often be preserved at critical points after truncation if the symmetries of the model are preserved. Examples of universality can be found in studies of quantum spin models \cite{PhysRevLett.72.2777,sachdev2000}. The idea of universality was emphasized recently in the context of studying sigma models using quantum computers in  Ref.~\cite{PhysRevLett.123.090501}. The procedure of constructing a $n$-qubit lattice Hamiltonian for studying a QFT can be viewed as an extra regularization necessary for quantum computation and was referred to as ``qubit regularization'' of the QFT in Refs.~\cite{Singh:2019uwd, Singh:2019jog}.

As with any form of regularization, a procedure to define the continuum limit of the $n$-qubit model is necessary. How this limit emerges is not obvious in the $n$-qubit model, at least when $n$ remains finite. A common technique followed by many groups is to approach the continuum limit using the na\"\i{}ve procedure of making $n$ large, which takes us back to the traditional theory \cite{Raychowdhury:2018osk,Klco:2019evd,Raychowdhury:2019iki,Bender:2020ztu,Davoudi:2020yln}. 
An interesting unanswered theoretical question is whether all QFTs can in principle be obtained using suitable continuum limits of $n$-qubit models where $n$ remains finite and, if so, what is the minimal value of $n$ necessary for each QFT? 

\begin{figure}
\includegraphics[width=0.9\linewidth]{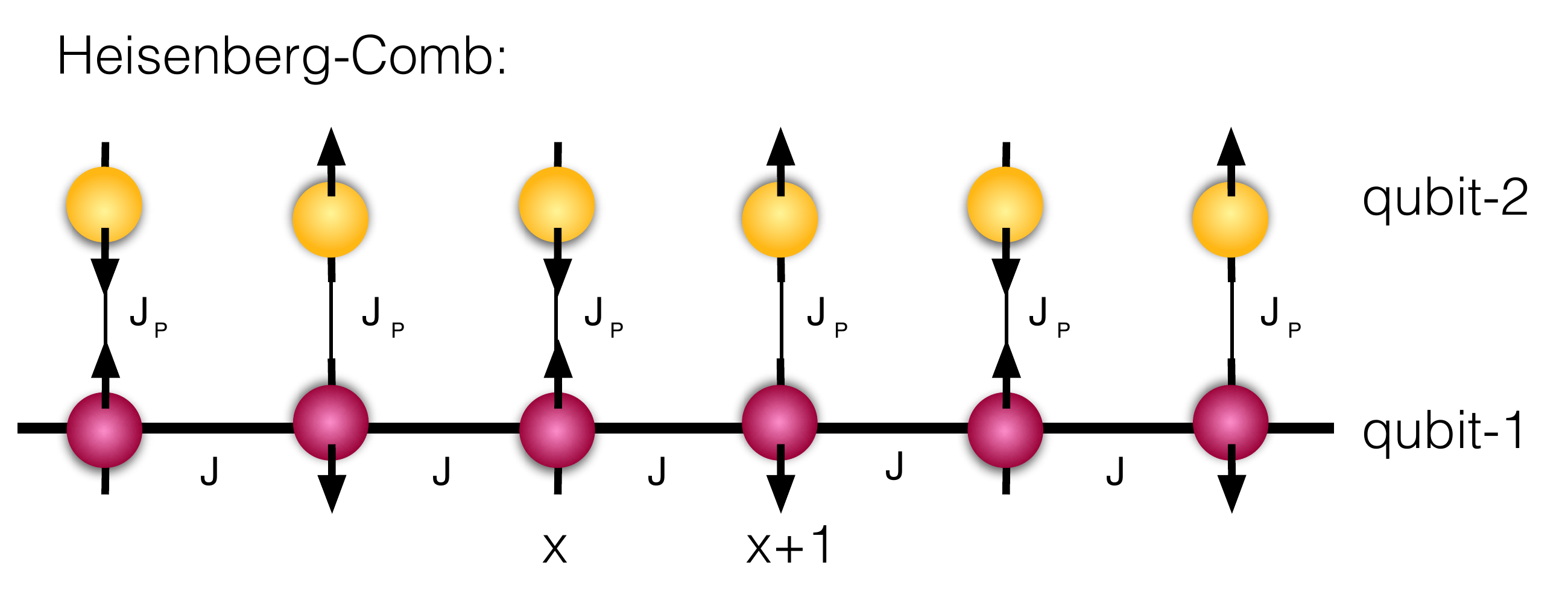}\hskip0.3in
\caption{A pictorial representation of the Heisenberg comb whose Hamiltonian is given in \cref{eq:hcombH}. The asymptotically free QFT described by \cref{eq:nlm} is reproduced in the limit $J/J_p \rightarrow \infty$.}
\label{fig:model}
\end{figure}

In this Letter, we show that the asymptotically free $1+1$-dimensional $O(3)$ nonlinear sigma model, described in Euclidean time $\tau$ by the continuum action
\begin{align}
  S[ \mathbf{n} ] = \frac{1}{2g^2} \int d\tau dx\ 
  \partial_\mu\mathbf{n} \cdot \partial_\mu\mathbf{n}\,,
  \label{eq:nlm}
\end{align}
with \(\mathbf{n}(x, \tau) \in O(3)\), 
can be regularized successfully using the $2$-qubit-per-site Hamiltonian
\begin{align}
H = \sum_i \quad J_p \ H_{(i,1),(i,2)} + J \ H_{(i,1),(i+1,1)},
\label{eq:hcombH}
\end{align}
which we illustrate pictorially in \cref{fig:model} and refer to as the Heisenberg comb. The continuum limit emerges when $J/J_p \rightarrow \infty$.
Note that $\mathbf{n}(x,\tau)$ in \cref{eq:nlm} is a  classical 3-vector field of unit magnitude, while $H_{(i,a),(j,b)} = \mathbf{S}_{i,a} \cdot \mathbf{S}_{j,b}$ is the standard Heisenberg interaction between spin-half operators $\mathbf{S}_{i,a}$ and $\mathbf{S}_{j,b}$, where $i,j$ label one-dimensional spatial lattice sites and $a,b=1,2$ label the 2-qubit spaces.

The idea of qubit regularization of field theories is not new and was introduced many years ago in the $D$-theory formulation of field theories \cite{Brower:1997ha,Brower:2003vy}. The challenge is to search the space of all $n$-qubit lattice Hamiltonians in $d$ spatial dimensions to discover the correct quantum critical point where the original continuum quantum field theory in $d+1$ space-time dimensions is recovered. The fine tuning to the quantum critical point would then naturally define the procedure to obtain the continuum limit of the $n$-qubit model. Qubit regularization of conformal field theories that emerge naturally at second order quantum critical points between two different phases are easy to construct with finite values of $n$. In this case, one has to just preserve the important symmetries of the original QFT and tune a single relevant parameter to the correct quantum critical point. While this technique is well known in the literature, the approach has found new applications recently \cite{Banerjee:2019jpw}.

In contrast, qubit regularization of asymptotically free theories like QCD and the nonlinear sigma model given in \cref{eq:nlm} is much more challenging, especially when $n$ is finite and fixed. This is the central topic of our Letter. In this case, in addition to preserving the symmetries of the QFT, one has to discover the critical point with the correct marginally relevant coupling that preserves the physics at all length scales from the infrared (IR) to the ultraviolet (UV). The IR is usually characterized by a physical correlation length $\xi$, while the UV is characterized by a minimum lattice size $L_{\rm min} \ll \xi$ at which the universal physics of the QFT can be observed using the lattice theory, where even $L_{\rm min}$ is much larger than the lattice spacing. 
In the $D$-theory formulation, this marginal operator is obtained through the size of an extra spatial dimension \cite{Wiese:2006kp}. It has been shown that asymptotic freedom in the two-dimensional $CP(N-1)$ models can be reproduced when the size of this extra dimension grows \cite{Beard:2004jr,Evans:2018njs}. Since the size of the extra dimension naturally increases the number of qubits per spatial lattice site, one can say that asymptotic freedom in the traditional $D$-theory approach can be obtained if $n$ is allowed to grow.\looseness-1

In this Letter, we explore whether asymptotic freedom may be achieved even with a fixed value of $n$ by discovering the correct marginally relevant coupling not related to an extra dimension. We show this explicitly in the case of the asymptotically free two-dimensional nonlinear $O(3)$ QFT described by \cref{eq:nlm}. Traditionally, this theory is regularized using the lattice Euclidean action
\begin{align}
S = -\kappa \sum_{\langle i,j\rangle,a} \phi^a_i \cdot \phi^a_j,
\label{eq:tradmodel}
\end{align}
where $i,j$ now label space-time lattice sites on a square lattice and $\phi^a_i$ ($a=1,2,3$) are the three components of a unit vector associated with the lattice site $i$. The continuum limit is obtained when $\kappa \rightarrow \infty$. Here we will argue that the same continuum physics can also be obtained using the Heisenberg-comb Hamiltonian discussed above.

The question of whether $2$-qubit models can reproduce the physics of the traditional model has been partially explored previously. The first exploration was performed in traditional lattice field theory using a Nienhuis-type action that can be viewed as a space-time lattice formulation of a 2-qubit Hamiltonian~\cite{Niedermayer:2016hzw}. Evidence was provided that a step-scaling function similar to that for the traditional model is reproduced at two different renormalized couplings. However, it was suggested that the continuum limit may not be reachable using such an approach. More recently, a $2$-qubit Hamiltonian was constructed by truncating the traditional infinite Hilbert space Hamiltonian onto the $2$-qubit subspace \cite{PhysRevD.99.074501}. The mass gap of the model was computed using tensor-network methods. While in the traditional model the mass gap would vanish as the bare coupling was lowered, i.e.,  $\kappa \rightarrow \infty$, the authors found that the mass gap did not vanish as the bare coupling was lowered in the $2$-qubit model. The authors also showed that, as the number of qubits per lattice site was increased, the mass gap quickly reduced, suggesting that more qubits will be necessary to recover the original theory. It is important to note that both of the above studies focused on a very specific class of Hamiltonians motivated by the traditional Hamiltonian. A more systematic search of the model space was never carried out.

Since the $SO(3)\subset O(3)$ symmetry plays an important role in the physics, we can narrow the search to the space of $2$-qubit models invariant under this symmetry. The phase diagram of these models is quite rich with several phases and quantum critical points separating them~\cite{Binder:2020nmm}. In particular, there are at least five distinct phases: phase A, where the two qubits form local spin singlets and the spin-triplet excitations are massive; phase B, where the spin triplets dominate and form a ferromagnet; phase C, where the spin triplets on neighboring sites form singlets and break translation invariance spontaneously; phase D, where spin triplets form a massive topological phase also referred to as the Haldane phase \cite{Haldane:1982rj,Haldane:1983ru,PhysRevLett.59.799,PhysRevB.81.064439}; and phase E where the long distance physics is a critical gapless phase described by level-one $SU(3)$ Wess-Zumino-Witten (WZW) conformal field theory~\cite{PhysRevB.55.8295}. Such a rich phase structure already suggests that previous studies could have missed the asymptotically free fixed point. \looseness-1

\begin{figure}
\includegraphics[width=0.9\linewidth]{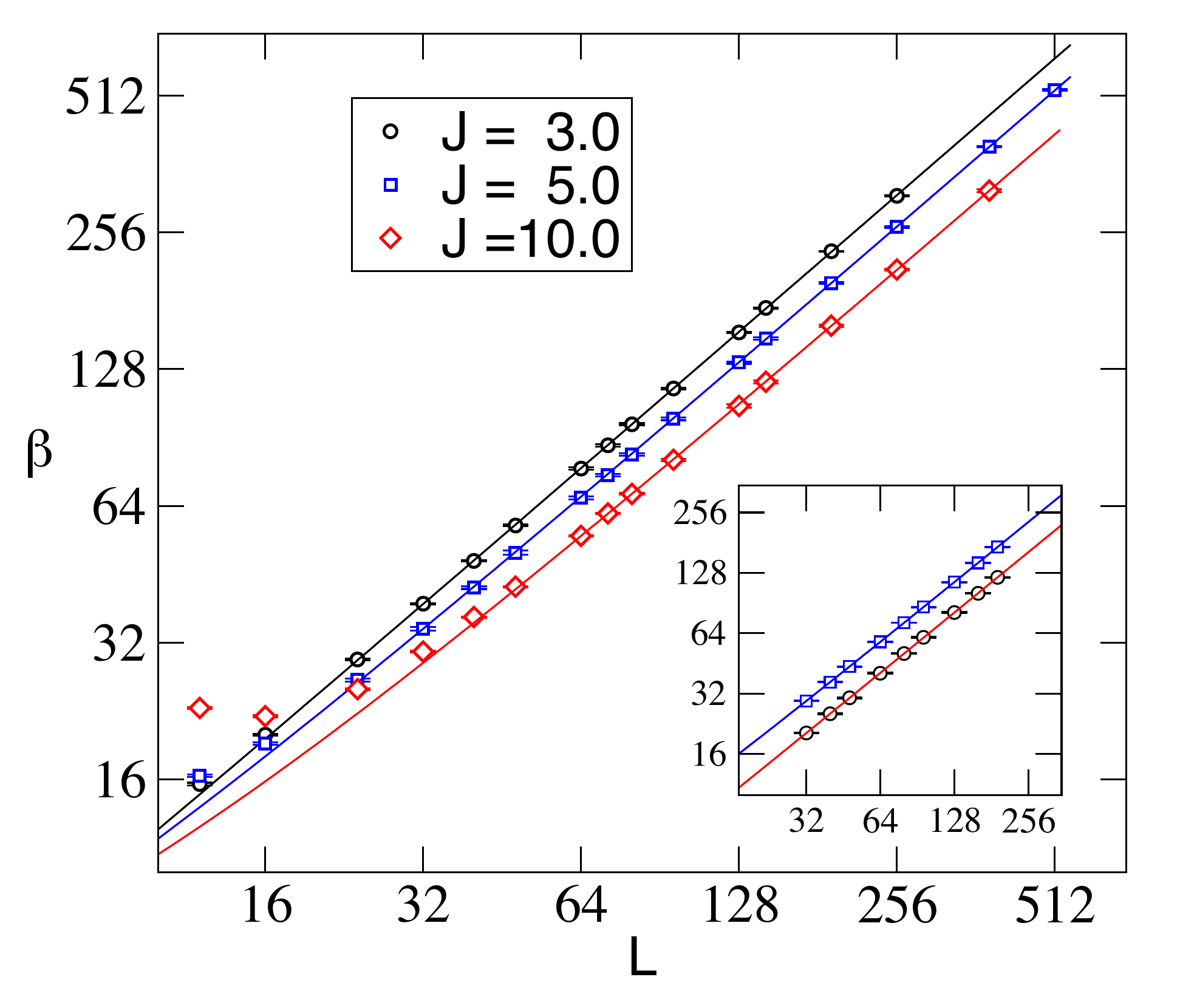}\hskip0.3in
\caption{The value of $\beta$ chosen for each value of $L$ during calculations of $\xi(L)$ in the Heisenberg comb at various values of $J$. The inset shows a similar plot for the symmetric ladder (circles) and asymmetric ladder (squares). While $\beta(L)$ depends on $J$ in the Heisenberg comb, it does not depend on $J_p$ for the parameters we have explored in the symmetric and asymmetric ladders.}
\label{fig:bvsL}
\end{figure}

\begin{figure*}
\includegraphics[width=0.9\linewidth]{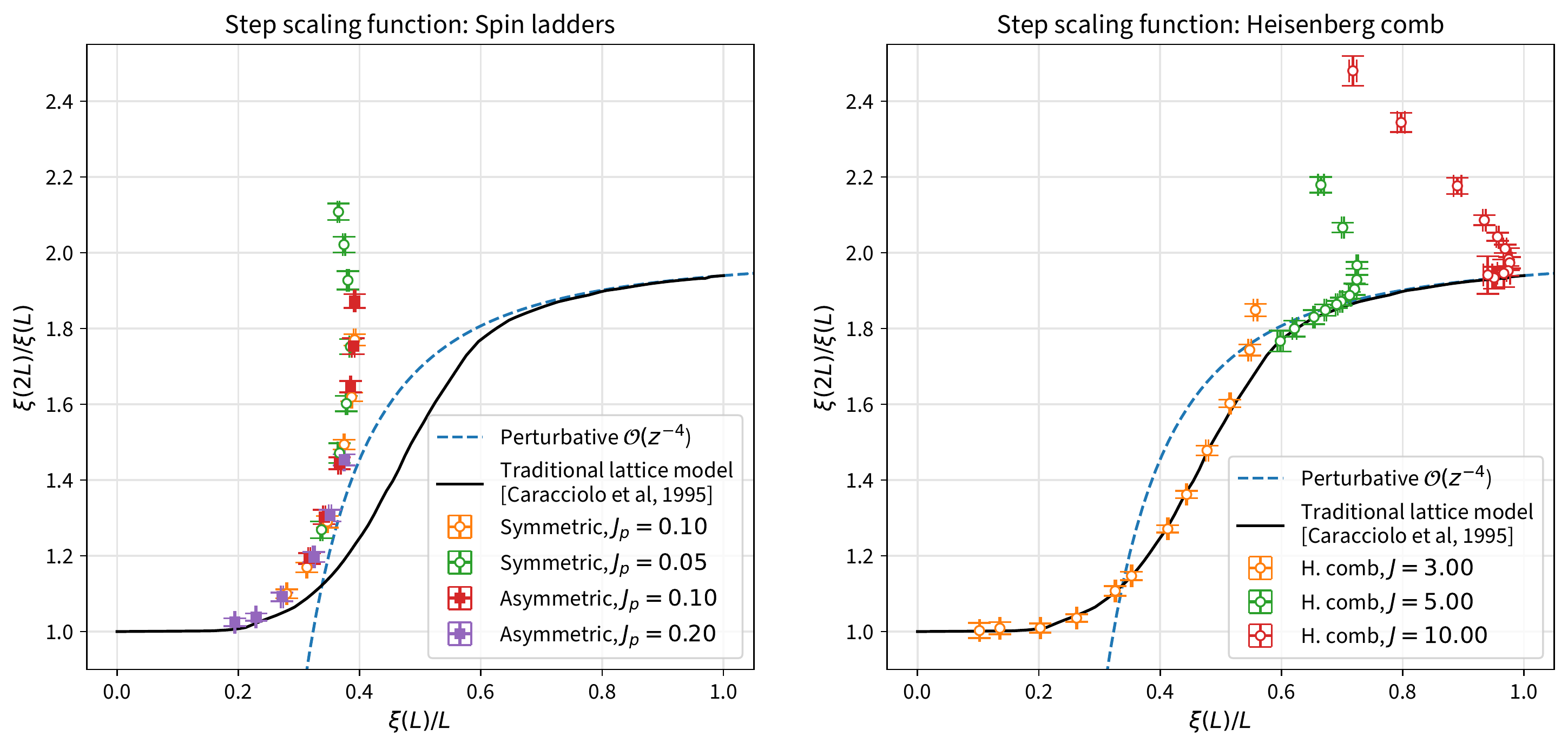}
\caption{Universal step-scaling function $F(z)=\xi(2L)/\xi(L)$ as a function of $z=\xi(L)/L$. The dark line in both plots is this function reproduced from \cite{PhysRevLett.75.1891}, where it was calculated using the traditional model defined by \cref{eq:tradmodel}. The dashed line is the function $F(z) = 2(1 - 0.0276/z^2 - 0.00258/z^4)$, computed in Ref.~\cite{PhysRevLett.75.1891} using perturbation theory near the asymptotically free fixed point. The left plot shows results for (1) a symmetric ladder ($J_1=J_2=1$; $J_p=0.10,0.20$) and (2) an asymmetric ladder ($J_1=1$, $J_2=0.5$; $J_p=0.05,0.10$). 
The right plot shows results for the Heisenberg comb at $J_1=J=3,5,10$; $J_2=0$;$J_p=1$. 
Each data point is constructed using two calculations, one at lattice size $L$ and another at $2L$, with $12 \leq L \leq 256$. }.
\label{fig:uf}
\end{figure*}

One way to parameterize the space of $2$-qubit models is to begin with spin-half ladders whose Hamiltonian takes the form  
\begin{align}
H \ =\  \sum_{x} \ J_p \ H_{(x,1),(x,2)} 
\ +\  & J_1 \ H_{(x,1),(x+1,1)} \nonumber \\
& \ +\ J_2 \ H_{(x,2),(x+1,2)},
\label{eq:spinl}
\end{align}
with three tunable couplings. When $J_p < 0$ and large compared to $J_1$ and $J_2$, we can access the physics of spin-1 chains. In these chains, when $J_1,J_2 < 0$, we obtain the ferromagnetic phase B. On the other hand, when $J_1,J_2 > 0$, we can access the physics of antiferromagnetic spin-1 chains, which is the starting point to accessing phases C, D and E. 
When $J_p=0$, it is well known that each of the two decoupled spin-half chains describe the long distance physics of 
\begin{align}
  S[ \mathbf n ] = \int d\tau dx\ 
  \Big\{\frac{1}{2g^2} \partial_\mu\mathbf{n} \cdot \partial_\mu\mathbf{n} +  \frac{i\theta}{4\pi} \mathbf{n}\cdot (\partial_\tau \mathbf{n} \times \partial_x \mathbf{n}) \Big\}
  \label{eq:nlmtheta}
\end{align}
at $\theta=\pi$ \cite{HaldanePLA93}. 
This theory is known to be critical and described by the $k=1$ $SU(2)$ WZW theory \cite{Affleck:1987ch,Affleck:1988px,Shankar:1989ee}. When $J_p > 0$,  $\theta$ is constrained to be zero, and the IR physics of \cref{eq:nlm} is reproduced \cite{shelton_antiferromagnetic_1996}. So it is likely that there exists a critical point in the three parameter space of $(J_1,J_2,J_p)$ with $J_p > 0$ that reproduces both the UV and IR physics of \cref{eq:nlm} correctly. In particular three critical points seem interesting candidates to explore: (1) the symmetric ladder where $J_1 = J_2 = 1$ and $J_p \rightarrow 0^+$, (2) the asymmetric ladder $J_1 > J_2$ and $J_p \rightarrow 0^+$, and (3) the Heisenberg comb $J_2 = 0$, $J_p = 1$, and $J_1 = J \rightarrow \infty$. We will show below that the Heisenberg comb is the correct $2$-qubit model.

In order to study whether the traditional model and the $2$-qubit model quantitatively reproduce the same physics in the continuum limit, we need to match the two theories at all physical scales from the IR to the UV. Asymptotically free theories are massive, and the correlation length $\xi$ defined by the mass gap sets the natural IR length scale. In order to probe the UV physics we put the system in a small box of physical size $L \ll \xi$. In fact, one can use a suitably defined finite size correlation length $\xi(L)$ even in the UV such that $\xi(L\rightarrow \infty) = \xi$. One definition of such a length scale is the second moment definition \cite{Caracciolo:1992nh}, $\xi(L) = [\sqrt{(G_0/G_1)-1} ]/[2\sin(\pi/L)]$
where 
\begin{align}
G_k = \sum_{(x,t)} 
\big\langle \phi^3_{(x,t)}\ \phi^3_{(0,0)}\big\rangle \ e^{i 2\pi k x/L}.
\label{eq:corrk}
\end{align}
A quantitative way to probe all physical scales from the UV to the IR using $\xi(L)$ is the universal step-scaling function $F(z) = \xi(2L)/\xi(L)$, where $z = \xi(L)/L$. 
This function is a signature of the asymptotically free QFT. It probes the IR physics for $z\rightarrow 0$ and the UV physics for $z\rightarrow \infty$, where it can computed using perturbation theory.
For the traditional lattice model defined in \cref{eq:tradmodel}, $F(z)$ was computed long ago \cite{PhysRevLett.75.1891}. Here we compute $F(z)$ for the spin-ladder models defined in \cref{eq:spinl} using well established continuous time Monte Carlo methods  \cite{PhysRevLett.77.5130,PhysRevB.59.R14157,RevModPhys.83.349}. 
In particular, there is no sign problem for the models we study here \cite{Evertz:2000rk}.

The calculations described focus on a few specific points in the phase space of spin ladders. In particular, we consider the symmetric ladder with $J_1=J_2=1$ at $J_p = 0.20$, $0.10$, the asymmetric ladder with $J_1=1, J_2=0.5$ at $J_p = 0.10, 0.05$, and the Heisenberg comb with $J_2=0, J_p=1$, at $J_1= J = 3,5,10$. To compute $\xi(L)$, we replace  \cref{eq:corrk} with
\begin{align}
G_k = \frac{1}{Z}\int d\tau \sum_x 
\mathrm{Tr}\Big[{\cal O}(x,\tau) {\cal O}(0,0)e^{-\beta H}\Big]\ e^{i 2\pi k x/L},
\label{eq:qcorrk}
\end{align}
where ${\cal O}(x,\tau) = e^{\tau H} [(-1)^x(S^z_{x,1}-S^z_{x,2})] e^{-\tau H}$ is the usual Heisenberg operator in imaginary time $\tau$, $Z = \mathrm{Tr}(e^{-\beta H})$ is the thermal partition function, and $\beta$ is the inverse temperature. We note that, on each lattice site, ${\cal O}$ creates the z component of the triplet from the singlet and vice versa. The $(-1)^x$ is required to capture the antiferromagnetic nature of the long distance physics. We study lattice sizes in the range $12 \leq L \leq 512$.
In a Hamiltonian formulation, spatial correlation functions, \cref{eq:qcorrk}, will, in general, be different from temporal correlation functions:
\begin{align}
\tilde{G}_k = \frac{1}{Z}\int d\tau \sum_x 
\mathrm{Tr}\Big[{\cal O}(x,\tau) {\cal O}(0,0)e^{-\beta H}\Big]\ e^{i 2\pi k \tau/\beta},
\label{eq:qcorrkt}
\end{align}
where, as in the traditional model, \cref{eq:tradmodel}, these two are identical on square lattices due to space-time rotational symmetry. In our calculations, we tune $\beta$ as a function of $L$ to make $G_k \approx \tilde{G}_k$. These fine-tuned values of $\beta$ are plotted as a function of $L$ in \cref{fig:bvsL}, which shows that for each of our studies $\beta/L$ becomes a constant for large $L$, as expected in a relativistic theory.

Our results for $F(z)$ in the qubit models are shown in two plots in \cref{fig:uf}, along with results from the traditional model recreated from \cite{PhysRevLett.75.1891}. In the left plot, we show our results for the symmetric ladder and the asymmetric ladder. These results show that neither of these models reproduce the traditional model in the UV, although they are $SO(3)$ symmetric and massive in the IR. Thus, the spin ladders may be described by \cref{eq:nlm} in the IR \cite{shelton_antiferromagnetic_1996}, but in the UV they are most likely described by two decoupled $k=1$ $SU(2)$ WZW conformal field theories, as one might expect. In fact, for small values of $J_p$ the mass gap in the symmetric case is known to increase linearly as $0.41(1) J_p$ implying that $J_p$ is not the marginally relevant coupling we are looking for \cite{greven_monte_1996}. 

In contrast to spin ladders, when $F(z)$ is computed in the Heisenberg comb, it matches the traditional model well for all values of $L > L_{\rm min}$. When $J=\num{3},\num{5},\num{10}$, we find that, in lattice units, $L_{\rm min} \approx 30,100,400$ and $\xi \approx \num{25},\num{600},\num{200000}$, respectively. We observe that the UV scale $L_{\rm min}$ increases with $J$, 
and in the limit $J \rightarrow \infty$ the asymptotically free critical point is recovered. From Wilson's renormalization group perspective, after blocking to a scale of $L_{\rm min}$, the $2$-qubit model turns into an $n$-qubit model where $n=2 L_{\rm min}$. The traditional infinite Hilbert space of the continuum asymptotically free fixed point is recovered in the $J\rightarrow \infty$ limit.

Note that we define the physics of the Heisenberg comb by setting $J_p=1$, $J_1=J$ and study the limit $J \rightarrow \infty$. We could have instead set $J_1=1$ and $J_p = 1/J$ and obtained the same physics. However, this would force us to study asymmetric lattices since now the values of $\beta$ in \cref{fig:bvsL} would need to be multiplied by $J$. One way to understand our choice is to recognize that the value of $J_p$ sets a UV energy scale for our problem. The physics of \cref{eq:nlm} only emerges at temperatures much smaller than $J_p$. So it is natural in the qubit regularization to set $J_p = 1$ and study the physics at large values of $J_1=J$. 

\begin{figure}
    \centering
    \includegraphics[width=\linewidth]{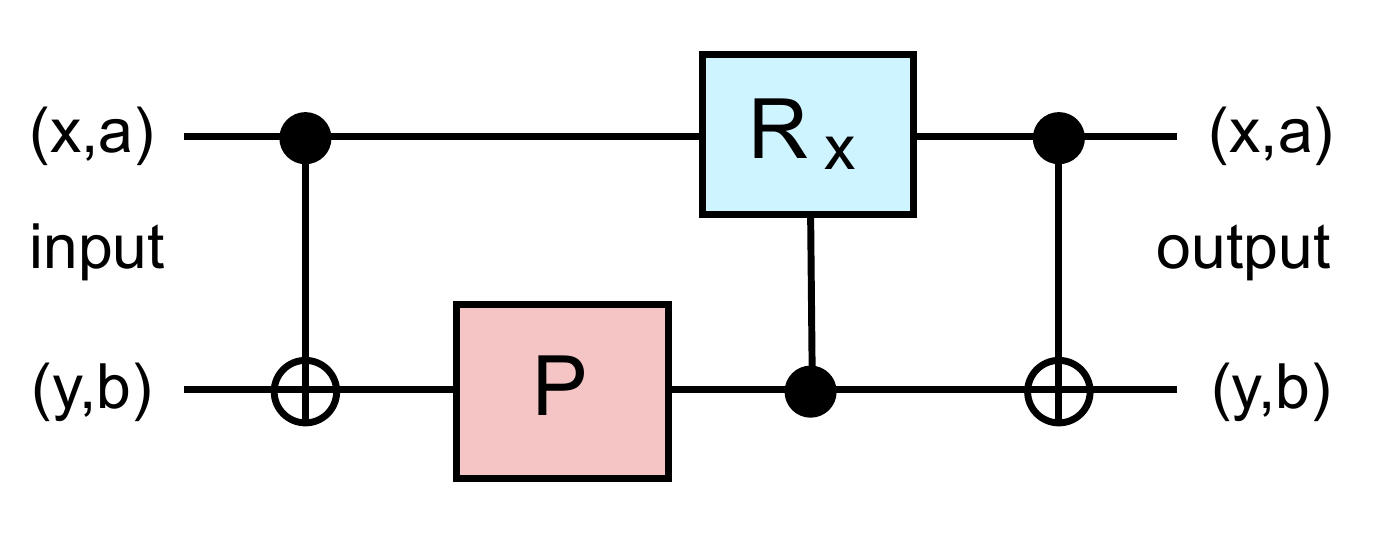}
    \caption{Quantum circuit implementing $e^{-iJ (H_{(x,a),(x,b)}+1/4) t}$ on two qubits $(x,a)$ and $(y,b)$ with exactly three entangling gates and one single-qubit gate. The first and second lines represent the qubits $(x,a)$ and $(y,b)$. The phase rotation $P$ and the $X$-rotation $R_X(\phi)$ are defined in \cref{eq:Xrot}. The angle $\phi = J_p t/2$ for terms in $H_1$ and $\phi = J t/2$ for terms in $H_2$ and $H_3$.}
    \label{fig:circuit}
\end{figure}

An important motivation for discovering a qubit regularization with a finite $n$ is the ability to study the real time evolution of the QFT on a quantum computer. The simplicity of the Heisenberg-comb Hamiltonian allows us to implement the Trotterized time-evolution operator of the theory using a short-depth quantum circuit. To construct this, we write the Hamiltonian as a sum of three commuting terms: $H = J_p H_1 + J (H_2 + H_3)$ where $
H_1 = \sum_{i\in x} H_{(x,1),(x,2)}$,
$H_2 = \sum_{i\in x_e} H_{(x_e,1),(x_e+1,1)}$, and $H_3 = \sum_{i\in x_o} H_{(x_o,1),(x_o+1,1)}$, where $x_e$ are even sites and $x_o$ are odd sites. Using the standard Trotter approach, we can then write
\begin{equation}
    e^{-iHt} = e^{-iJ_p H_1 t}e^{-iJ H_2 t}e^{-iJ H_3 t} \ + \ \mathcal{O}(t^2).
\end{equation}
In the computational basis, that is, the $\ket{\pm}$ basis for each spin, we can view the Hamiltonian $H_{(x,a),(y,b)}$ as a 2-qubit operator whose matrix elements are given by a $4\times 4$ matrix.
The corresponding time-evolution operator $e^{-iJ [H_{(x,a),(y,b)}+1/4] t}$ (with an additional global phase that can be easily undone) takes the form
\begin{equation}
\label{eq:unitary}
    \begin{bmatrix}
    e^{-iJt/2} & 0 & 0 & 0 \\
    0 & \cos{(J t/2)} & -i\sin{(J t/2)} & 0 \\
    0 & -i\sin{(J t/2)} & \cos{(J t/2)} & 0 \\
    0 & 0 & 0 & e^{-iJt/2} \\
    \end{bmatrix}.
\end{equation}
This unitary transformation can be implemented using two controlled NOT gates, one controlled unitary gate that implements the $X$-rotation $R_X(\phi)$ and one single-qubit phase rotation $P(\phi)$, given by
\begin{align}
R_X(\phi)=\ 
    \begin{bmatrix}
    \cos\phi & -i\sin\phi\\
    -i\sin\phi & \cos\phi\\
    \end{bmatrix}, \ \ \  P(\phi) = \protect\begin{pmatrix} 
e^{-i\phi} & 0\\ 0 & 1 
\protect\end{pmatrix}.
\label{eq:Xrot}
\end{align}
The quantum circuit that implements the full unitary transformation in \cref{eq:unitary} is given in \cref{fig:circuit}. The simplicity of this circuit suggests that near-term quantum computers may suffice to simulate dynamics for short times. An interesting first step is to show that the critical point in question is asymptotically free using a quantum simulator. Similar experiments have been done for other models on a variety of platforms \cite{zhang2017observation,xu2020probing,keesling2019quantum}.

The interesting question for the quantum computation of asymptotically free theories like QCD is whether it is better to regulate the theory using a small number of qubits per lattice site and show that asymptotic freedom emerges dynamically in the usual continuum limit, or use a formulation with a large number of qubits per lattice site that approximates the classical model as is normally done. This work on the $O(3)$ model demonstrates that asymptotic freedom \emph{does not necessarily require} an infinite dimensional local Hilbert space in a lattice model, although intuition might lead us to believe this is necessary. By suitably adjusting the lattice size $L$ as $J$ is increased to approach the continuum limit, one can stay on the universal scaling curve and thus deduce the properties of the continuum theory. The question remains: which approach is more efficient to implement on a quantum computer? Here we have shown that the quantum circuit for the Heisenberg comb is simple, but whether the growth of complexity with lattice size
eventually makes the other approach more efficient remains to be investigated.\looseness-1

S.C. and H.S. would like to thank Thomas Barthel, Moritz Binder, Ribhu Kaul, Hanqing Liu, and Uwe-Jens Wiese for helpful discussions on the subject. 
The material presented here is based on work supported by the U.S. Department of Energy, Office of Science --- High Energy Physics Contract KA2401032 (Triad National Security, LLC Contract Grant No. 89233218CNA000001) to Los Alamos National Laboratory. 
S.C. is supported by a Duke subcontract of this grant. 
S.C. and H.S. were also supported for this work in part by the U.S. Department of Energy, Office of Science, Nuclear Physics program under Award No. DE-FG02-05ER41368.

\bibliographystyle{apsrev4-2}
\bibliography{ref}

\end{document}